\documentclass[12pt]{article}
\usepackage{amsmath,amssymb}
\usepackage{a4}
\usepackage{epsfig}

\begin{document}

\title{Space-time noncommutativity with a bifermionic parameter}
\author{D.~M.~Gitman\thanks{E.mail: gitman@dfn.if.usp.br} 
and D.~V.~Vassilevich\thanks{On leave from V.~A.~Fock Institute of
Physics, St.~Petersburg University, Russia. E.mail: dmitry@dfn.if.usp.br} \\
\textit{Instituto de F\'isica, Universidade de S\~ao Paulo,}\\
\textit{Caixa Postal 66318 CEP 05315-970, S\~ao Paulo, S.P., Brazil}}
\maketitle

\begin{abstract}
We consider a Moyal plane and propose to make the noncommutativity
parameter $\Theta^{\mu\nu}$ bifermionic, i.e.,  composed of two
fermionic (Grassmann odd) parameters. The Moyal product then contains
a finite number of derivatives, which allows to avoid difficulties
of the standard approach. As an example,
we construct a two-dimensional noncommutative
field theory model based on the Moyal product with a bifermionic parameter
and show that it has a
locally conserved energy-momentum tensor. The model has no problems
with the canonical quantization and appears to be renormalizable.
\end{abstract}

Noncommutativity enters field theory from many sides \cite{NCrevs},
including string theory and Quantum Hall Effect. In \cite{Doplicher:1994zv}
it was shown in particular that a combination of quantum mechanics and
classical gravity leads to uncertainty relations between space-time
coordinates, so that these coordinates become noncommutative (NC). It is
rather possible that on a more fundamental level of a fully quantized field
theory the NC parameter is \emph{Grassmann even and bifermionic}, i.e., is a
composite of two fermionic (Grassmann odd) parameters. This possibility will
be studied in this paper with a simple example of two dimensional quantum
field theory. We show that having a bifermionic NC parameter helps to
overcome, or at least to by-pass, many difficulties of the standard approach
to NC field theories.

Previously, a bifermionic NC parameter appeared in the context of a
construction of Lorentz covariant supersymmetric brackets \cite{Zh}.
However, consequences of such NC parameter for quantum field theory
were not discussed.

It is well known that
the standard Moyal product is non-local. As a consequence, field theories
based on this product are also non-local. Therefore, many essential
properties of local theories disappear in their NC counterparts. For
example, it is hard to define a locally conserved energy-momentum tensor %
\cite{Gerhold:2000ik}. Even more problems appear in space-time NC theories.
The presence on an infinite number of time derivatives makes the standard
canonical Hamiltonian approach hardly possible. One can instead develop and
Ostrogradski-type formalism \cite{Ostr} (see also \cite{Rosen}) at the
expanse of introducing an a new evolution parameter in addition to the usual
time coordinate. Unfortunately, the canonical field operators then depend on
this additional parameter thus preventing us from using the full power of
canonical methods. Note, that one can define a Poisson structure which is
suitable for analyzing gauge symmetries in space-time NC theories by looking
at their constraint algebra \cite{Vassilevich:2005fk}. Although this method
is rather useful in working with particular models \cite{Vassilevich:2006uv}%
, it is hard to say whether it can be applied to quantization. As a
consequence of these difficulties with constructing a canonical Hamiltonian
formalism, the unitarity in space-time noncommutative theories is not
guaranteed, see \cite{unitarity}. There is a rather radical approach
to NC theories, based on twisted quantization rules, which avoids
the problems mentioned above, but also removes practically all
effects of the noncommutativity (see \cite{Fiore:2007vg} and references
therein).

We suggest the following geometry. Consider a two-dimensional plane 
with the coordinates being Grassmann even elements of
Berezin algebra (see \cite{GitTy90})  (bosonic variables of pseudoclassical
mechanics) and a star product such that $x^{\mu }\star x^{\nu
}-x^{\nu }\star x^{\mu }=i\Theta ^{\mu \nu }$, and $\Theta ^{\mu \nu }$
is a real Grassmann even constant. Note, that in two dimensions this inevitably
means that space and time do not commute. We take the simplest possible
bifermionic representation for $\Theta ^{\mu \nu }$, namely 
\begin{equation}
\Theta ^{\mu \nu }=i\theta ^{\mu }\theta ^{\nu },  \label{Theta}
\end{equation}%
where $\theta ^{\mu }$ are constant real fermions (Grassmann odd constants),
$\theta^\mu\theta^\nu = -\theta^\nu\theta^\mu$.
Bifermionic constants appear naturally in pseudoclassical models of
relativistic particles with spin, see e.g. \cite{Gitman:1994rt,GGT}. In such
models these constants become numbers upon quantization\footnote{
For completeness we note that there is an alternative approach \cite{plu}
based on introduction of additional fields, which leads, however, to
equivalent results, see \cite{Gitman:2002xa} for a discussion.}.

The usual Moyal product with the NC parameter given by (\ref{Theta})
terminates at the second term due to the property $(\theta^\mu)^2=0$,
\begin{eqnarray}
f_1\star f_2&=& \exp \left( \frac i2 \Theta^{\mu\nu} \partial_\mu^x
\partial_\nu^y \right)f_1(x)f_2(y)\vert_{y=x}  \notag \\
&=&f_1\cdot f_2 -\frac 12 \theta^0\theta^1 (\partial_0 f_1
\partial_1f_2-\partial_1f_1\partial_0f_2).  \label{bifstar}
\end{eqnarray}
Symmetrized star-product therefore equals to the ordinary one. In
particular, $f\star f=f\cdot f$.

Let us construct a field theory model on the NC manifold we have just defined.
We take the action in the form 
\begin{eqnarray}
&&S=\int d^{2}x\left( \frac{1}{2}(\partial _{\mu }\varphi _{1})^{2}+\frac{1}{%
2}(\partial _{\mu }\varphi _{2})^{2}+\frac{1}{2}(\partial _{\mu }\varphi
)^{2}-\frac{1}{2}m_{1}^{2}\varphi _{1}^{2}-\frac{1}{2}m_{2}^{2}\varphi
_{2}^{2}-\frac{1}{2}m^{2}\varphi ^{2}\right.   \notag \\
&&\qquad \qquad \left. -ei[\varphi _{1},\varphi _{2}]_{\star }\star \varphi
\star \varphi -\lambda \varphi _{\star }^{n}\right) .  \label{act1}
\end{eqnarray}%
The signature of space-time is $(+-)$. $\varphi $, $\varphi _{1}$ and $%
\varphi _{2}$ are real scalar fields. Let us explain briefly where different
terms come from. We need at least two scalar fields, $\varphi _{1}$ and $%
\varphi _{2}$, to be able to construct a non-trivial star product $[\varphi
_{1},\varphi _{2}]_{\star }=\varphi _{1}\star \varphi _{2}-\varphi _{2}\star
\varphi _{1}$. The space-time integral of a star-commutator vanishes,
therefore to obtain a nonzero term in the action we have to multiply it by
another field, which we choose as $\varphi \star \varphi $. Star-commutator
of two real fields is imaginary, and we multiply our interaction term by an
imaginary coupling constant $ie$. We add standard kinetic and mass terms for
all fields and a self interaction term $\lambda \varphi _{\star }^{n}$ with $%
n$-th star-power of $\varphi $. The last term is needed to make the quantum
theory based on (\ref{act1}) non-trivial, see below.

By using (\ref{bifstar}) one can rewrite the interaction terms 
in (\ref{act1}) as 
\begin{equation}
S_{\mathrm{int}}=\int d^{2}x\left( 
ei\theta ^{0}\theta ^{1}(\partial _{0}\varphi
_{1}\partial _{1}\varphi _{2}-\partial _{1}\varphi _{1}\partial _{0}\varphi
_{2})\varphi ^{2}
-\lambda \varphi ^{n}\right) .  \label{interact}
\end{equation}%
We can also rewrite this term in a manifestly Lorentz-invariant form: 
\begin{equation}
S_{\mathrm{int}}=\int d^{2}x\left( -\frac{1}{2}ei\epsilon _{\rho \sigma
}\theta ^{\rho }\theta ^{\sigma }\epsilon ^{\mu \nu }\partial _{\mu }\varphi
_{1}\partial _{\nu }\varphi _{2}\, \varphi^2
-\lambda \varphi ^{n}\right) .
\label{interLor}
\end{equation}%
Note, that since $\epsilon _{\rho \sigma }$ is Lorentz invariant, one
must not transform the external parameters $\theta ^{\rho }$ in order to
achieve invariance of the action. This is a peculiar feature of
two-dimensional theories. In higher dimensions one should probably use the
twisted Poincare invariance like in NC theories with generic $\Theta $ \cite%
{twistP}.

The equations of motion for this model read: 
\begin{eqnarray}
&&-\Box \varphi - m^2\varphi -\lambda n \varphi^{n-1}
+2ie\theta^0\theta^1\varphi (\partial_0\varphi_1 \partial_1 \varphi_2
-\partial_1\varphi_1 \partial_0 \varphi_2)=0,  \notag \\
&&-\Box \varphi_1 - m_1^2\varphi_1 - 2ie\theta^0\theta^1\varphi
(\partial_0\varphi \partial_1\varphi_2 - \partial_0\varphi_2
\partial_1\varphi)=0,  \label{eom} \\
&&-\Box \varphi_2 - m_2^2\varphi_2 - 2ie\theta^0\theta^1\varphi
(\partial_0\varphi_1\partial_1\varphi -\partial_1\varphi_1
\partial_0\varphi)=0.  \notag
\end{eqnarray}

In NC theories it is problematic to define a locally conserved
energy-momentum tensor \cite{Gerhold:2000ik}. In our model no such problems
arise. Since the action (\ref{act1}) contains first derivatives at most, the
standard energy-momentum tensor 
\begin{equation}
T^\nu_{\ \mu}=\varphi_{;\mu}\frac{\delta \mathcal{L}}{\delta \varphi_{;\nu}}
-\delta^\nu_{\ \mu} \mathcal{L}  \label{Tmn}
\end{equation}
(where $\mathcal{L}$ is the Lagrangian density and $\varphi_{;\mu}\equiv
\partial_{\mu}\varphi$) is locally conserved. Indeed, by taking the
components 
\begin{eqnarray}
&&T^0_{\ 0}=\frac 12 ( (\partial_0\varphi)^2 + (\partial_1\varphi)^2 +
m^2\varphi^2 +(\partial_0\varphi_1)^2 + (\partial_1\varphi_1)^2 +
m_1^2\varphi_1^2  \notag \\
&&\qquad\quad +(\partial_0\varphi_2)^2 + (\partial_1\varphi_2)^2 +
m_2^2\varphi_2^2) +\lambda \varphi^n ,  \notag \\
&&T^1_{\ 1}=\frac 12 ( -(\partial_0\varphi)^2 - (\partial_1\varphi)^2 +
m^2\varphi^2 -(\partial_0\varphi_1)^2 - (\partial_1\varphi_1)^2 +
m_1^2\varphi_1^2  \notag \\
&&\qquad\quad -(\partial_0\varphi_2)^2 - (\partial_1\varphi_2)^2 +
m_2^2\varphi_2^2) +\lambda \varphi^n ,  \label{Tcomp} \\
&&T^0_{\ 1}=-T^1_{\ 0}=\partial_0\varphi \partial_1\varphi
+\partial_0\varphi_1 \partial_1\varphi_1 +\partial_0\varphi_2
\partial_1\varphi_2  \notag
\end{eqnarray}
and using the equations of motion (\ref{eom}) one can easily check the
conservation equation 
\begin{equation}
\partial_\mu T^\mu_{\ \nu}=0.  \label{consT}
\end{equation}

It is interesting to note that the energy-momentum tensor (\ref{Tcomp}) does
not contain the NC coupling $e\theta ^{0}\theta ^{1}$. This can be understood
looking at (\ref{interLor}). In curved-space generalizations the indices $%
\mu $, $\nu $ of the partial derivatives should be regarded as curved 
(world) ones,
so that $\epsilon ^{\mu \nu }$ becomes a tensor with the components $\pm
g^{-1/2}$ (with $g$ being determinant of the metric). If the indices of $%
\theta ^{0,1}$ are regarded as flat (tangential), 
then $\epsilon _{\rho \sigma }=\pm 1$,
the whole first term in (\ref{interLor}) does not depend on the metric and,
therefore, does not contribute to the metric stress-energy tensor.

Since the model contains first time derivatives at most, the Hamiltonian
analysis proceeds without any complications. The canonical momenta are
defined as variational derivatives of the action w.r.t.\ temporal
derivatives of corresponding fields 
\begin{eqnarray}
&&\pi =\partial _{0}\varphi ,  \notag \\
&&\pi _{1}=\partial _{0}\varphi _{1}+ei\theta ^{0}\theta ^{1}\,\partial
_{1}\varphi _{2}\,\varphi ^{2},  \label{momenta} \\
&&\pi _{2}=\partial _{0}\varphi _{2}-ei\theta ^{0}\theta ^{1}\,\partial
_{1}\varphi _{1}\,\varphi ^{2}.  \notag
\end{eqnarray}%
No constraints appear in this model.

The canonical Hamiltonian reads 
\begin{eqnarray}
&&H=\int dx(\pi \partial _{0}\varphi +\pi _{1}\partial _{0}\varphi _{1}+\pi
_{2}\partial _{0}\varphi _{2}-\mathcal{L})  \notag \\
&&\quad =\int dx\left( \frac{1}{2}(\pi ^{2}+(\partial _{1}\varphi
)^{2}+m^{2}\varphi ^{2}+\pi _{1}^{2}+(\partial _{1}\varphi
_{1})^{2}+m_{1}^{2}\varphi _{1}^{2}\right.   \label{Ham} \\
&&\qquad \left. +\pi _{2}^{2}+(\partial _{1}\varphi
_{2})^{2}+m_{2}^{2}\varphi _{2}^{2})+\lambda \varphi ^{n}+ei\theta ^{0}\theta
^{1}(-\pi _{1}\partial _{1}\varphi _{2}+\pi _{2}\partial _{1}\varphi
_{1})\varphi ^{2}\right) .  \notag
\end{eqnarray}%
Note, that after substituting in (\ref{Ham}) the canonical momenta in terms
of temporal derivatives of the fields by means of (\ref{momenta}) one
obtains the space integral of $T_{\ 0}^{0}$.

The canonical Hamiltonian is conserved by the construction, which can also
be checked by direct calculations.

Let us now proceed with the quantization. Since there are no constraints, we
simply replace the fields and their momenta by operators (which will be
denoted by the same letters with hats) that obey the canonical commutation
relations: 
\begin{equation}
\lbrack \hat{\varphi}(t,x),\hat{\pi}(t,x^{\prime })]=[\hat{\varphi}_{1}(t,x),%
\hat{\pi}_{1}(t,x^{\prime })]=[\hat{\varphi}_{2}(t,x),\hat{\pi}%
_{2}(t,x^{\prime })]=i\hbar \delta (x-x^{\prime }).  \label{ccom}
\end{equation}%
The Hamiltonian (\ref{Ham}) can be transformed into an operator without any
ordering ambiguities. The quantum Hamiltonian $\hat H$ 
obtained in this way is not
positive definite. However, the evolution operator constructed from $\hat H$
can be treated pertubatively w.r.t. the
charge $e$, and any perturbation series terminates at the first term. One
can then pass to the path integral formulation and define Feynman rules in a
perfectly standard way. On can think of $\theta ^{\mu }$ as of constant
fermion fields. Note, that in the contrast to \cite%
{Gitman:1994rt,Gitman:2002xa} we have no constraints here which fix the
value of the bifermionic parameter to a number. Upon quantization in the
effective action our fermions $\theta ^{\mu }$ are replaced by their averages
(precisely as usual fermionic fields). However, a trace of their Grassmann
nature remains: the perturbative series terminate. Therefore, in deriving
the Feynman diagrams from the path integral one can still consider $\theta
^{\mu }$ as a fermion (again, precisely as in the case of fields) and set
it to a number at the end of deriving expressions for quantum amplitudes.

\begin{figure}[htbp]
\begin{center}
\ \psfig{file=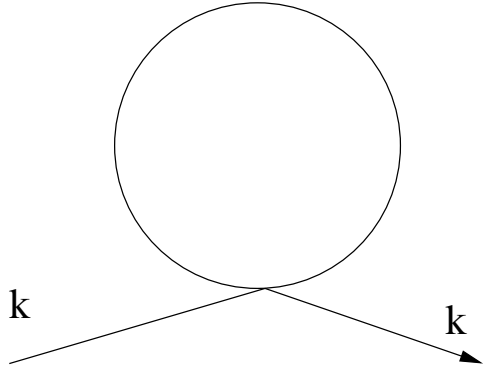}
\end{center}
\caption{An example of a
power-counting divergent diagram with the new interaction.
Left and right legs correspond to $\protect\varphi_1$ and $\protect\varphi_2$
respectively. The closed loop corresponds to $\protect\varphi$.}
\label{thefigure}
\end{figure}

Let us study renormalization of our model. Take $n=4$ in (\ref{act1}). If we
neglect for a while the interaction described by the charge $e$ we obtain a
model consisting of two free massive scalars ($\varphi _{1}$ and $\varphi
_{2}$) and another massive scalar with a $\varphi ^{4}$ selfinteraction.
This model is known to be multiplicatively renormalizable. Now, let us add
the missing term. Due to the fermionic nature of $\theta ^{0}$ and 
$\theta ^{1}$
the vertex generated by this interaction can enter a Feynman diagram at most
once. It is easy to see that the only possible one-particle irreducible
power-counting divergent diagrams with the charge $e$ are the ones
which have a pair of external $\varphi_1$, $\varphi_2$ legs and have no
external $\varphi$ legs, an example is given by Fig.~\ref{thefigure}. 
However, because of the structure $(\partial _{0}\varphi
_{1}\partial _{1}\varphi _{2}-\partial _{1}\varphi _{1}\partial _{0}\varphi
_{2})$ in the interaction term, such diagrams are proportional to $%
k_{0}k_{1}-k_{1}k_{0}=0$, i.e. they vanish identically. We
conclude, that noncommutativity does not bring any new divergences in the
model, so that it remains renormalizable. Note, that there are non-trivial
finite diagrams involving the noncommutative interaction. For example, by
using the $\varphi ^{4}$ vertex one can add an arbitrary number of pairs of
external $\varphi $ legs to the diagram of Fig.~\ref{thefigure}.

To summarize, we have suggested a space-time NC model with a bifermionic NC
parameter. In this model the Moyal product terminates at the second term,
and we have no problems caused by the nonlocality. In particular, this model
has a locally conserved energy-momentum tensor and a well defined
Hamiltonian. Also, we do not see any problem with quantization by the
canonical methods and by the path integral\footnote{%
Another example of a space-time NC theory having no problems with the path
integral to all order of perturbation theory is an integrable 2D gravity %
\cite{Vassilevich:2004ym}.}. Moreover, this model appears to be
renormalizable. An important lesson of our work is that noncommutativity
can lead to a rather mild modifications of the original commutative model.

What we have presented here is just a general idea which has to be
investigated further to put it on firmer grounds. For example, the unitarity
has to be checked explicitly. Although, due to the locality, there should be
no UV/IR mixing \cite{UVIR}, the IR behavior of our model has to be
explored. We used the simplest possible representation for the bifermionic
constant. Perhaps physically more preferable would be to represent it through
constant spinors (instead of constant vector fermions) as $\Theta ^{\mu \nu
}\propto \bar{\psi}[\gamma ^{\mu },\gamma ^{\nu }]\psi $. In two dimension
and for Majorana fermions this representation is essentially equivalent to (%
\ref{Theta}), but in higher dimensions it opens many new possibilities, in
particular, for constructing supersymmetric NC field models.

\textbf{Acknowledgement}. The work of D.M.G.\ was partially supported by
CNPq and FAPESP. D.V.V.\ was partially supported by the project
LSS.5538.2006.2.

\end{document}